\documentclass[11pt]{article}

\usepackage[preprint]{acl}

\usepackage{times}
\usepackage{latexsym}
\usepackage{booktabs}
\usepackage[T1]{fontenc}

\usepackage[utf8]{inputenc}

\usepackage{microtype}

\usepackage{inconsolata}

\usepackage{graphicx}
\usepackage{array} 
\usepackage{hyperref}
\usepackage[most]{tcolorbox}

\usepackage[hang,flushmargin]{footmisc}

\newenvironment{itemize*}%
  {\begin{itemize}%
    \setlength{\itemsep}{0.9pt}%
    \setlength{\parskip}{0.9pt}%
    \setlength{\topsep}{0.9pt}}%
  {\end{itemize}}

\usepackage[nameinlink]{cleveref}

\crefformat{section}{\S#2#1#3} 
\crefformat{subsection}{\S#2#1#3}
\crefformat{subsubsection}{\S#2#1#3}

\usepackage{todonotes}
\makeatletter
\newcommand*\iftodonotes{\if@todonotes@disabled\expandafter\@secondoftwo\else\expandafter\@firstoftwo\fi} 
\makeatother

\title{AInterviewer: A Platform for Designing and Conducting %
AI-led Qualitative Interviews}

\author{Tobias Priesholm Gårdhus \\
  University of Copenhagen \\
  \texttt{tpg@sodas.ku.dk} \\\And
  Nikolas Vitsakis \\
  IT University of Copenhagen \\
  \texttt{nvit@itu.dk} \\\And
  Fie Lejre Frederiksen \\
  University of Copenhagen \\
  \texttt{flf@sodas.ku.dk} \\\AND
  Anna Rogers \\
  IT University of Copenhagen \\
  \texttt{arog@itu.dk} \\\And
  Hjalmar Bang Carlsen \\
  University of Copenhagen \\
  \texttt{hc@sodas.ku.dk}
  }

\begin{document}
\maketitle

\begin{abstract}
There are now multiple proposals for systems based on Large Language Models (LLMs) to conduct automated qualitative interviews, but %
most of the current solutions rely on proprietary LLMs, which compromises reproducibility and data security. They also rely on LLMs for all interview tasks, which limits standardisation of question wording as well as control over question order. To address these issues, we introduce the \textit{AInterviewer}\footnote{Platform demo: \url{https://app.ainterviewer.dk}\\
YouTube demo: 
\url{https://www.youtube.com/watch?v=BOWuNsFVqAU}\\
Source code: \url{https://www.github.com/ainterviewer}
}platform, an open-source solution based on a multi-agent pipeline that combines controlled question administration of survey software with the flexibility of LLMs. \textit{AInterviewer} is an interdisciplinary effort designed to implement best practices of qualitative interviewing in social science, and it can run with locally hosted models to ensure security, transparency, and reproducibility. Our platform provides a web-based GUI supporting each phase of data collection: from interview guide design and pilot testing to interview distribution and data collection monitoring.
\end{abstract}

\section{Introduction to AI-interviewing}

The recent years saw multiple proposals for conversational AI tools to conduct interviews with humans \cite{chopra_conducting_2023, geiecke_conversations_2024, wuttke_ai_2024, xiao_tell_2020, jacobsen_chatbots_2025, sturgis_socbot_2025}. The so-called AI-interview systems rely on Large Language Models (LLMs) for asking open-ended and follow-up questions. This approach scales up qualitative interview techniques, which have traditionally been constrained by high costs of data collection \cite{chopra_conducting_2023, jacobsen_chatbots_2025}. Proponents argue that AI-led interviewing offers a new methodological middle ground between qualitative depth and quantitative scale, enabling researchers to collect large-N qualitative interview data in ways previously unattainable.

However, current research and the state of the art have largely relied on general-purpose commercial chatbots \cite{xiao_if_2020, chopra_conducting_2023}. While such setups allow researchers to conduct AI-led interviews with minimal preconfiguration, they have the following disadvantages: %
\begin{itemize*}
    \item limited control over question formulation, which can undermine comparability across interviews \cite{gerson_science_2020};
    \item lack of control over the model, which undermines reliability and reproducibility \cite{palmer_using_2024, staudinger2024reproducibility};
    \item insufficient control over data, which undermines qualitative requirements for confidentiality \cite{li2026agentic, shanmugarasa2025privacy, evertz2024whispers}.
\end{itemize*}

To establish AI-interviewing as a rigorous method, researchers need tools that combine conversational AI flexibility with controlled outputs and transparent system designs.

To address this gap, we created the \textit{AInterviewer}, a new platform for designing, instructing, testing, and distributing AI-interviews. Interviews are conducted by a multi-agent pipeline backed by open-sourced LLMs, where each agent handles different elements of the interview process. Unlike previous work, our approach enables fine-grained control of agentic behaviour through customizable interview guides, allowing for both deterministic ordering and responsive question formulation. For the deployment stage, our platform also includes built-in tools for pre-testing the interview, large-scale distribution, and experimental randomisation across and within interview conditions. To promote transparency, reproducibility, and collaborative scientific progress, \textit{AInterviewer} is released under the \textit{GNU Affero General Public License v3.0 (AGPLv3)}.

\section{Interview System Design}

\begin{figure}[!t]
    \centering
    \includegraphics[width=1\linewidth]{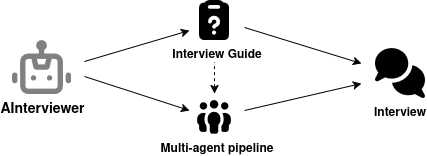}
    \caption{System design of the AInterviewer platform. The interview guide specifies the interview structure and how it may be adjusted in specific interviews based on information needs and prior conversational context.}
    \label{fig:simplified-agent-flow}
\end{figure}

\autoref{fig:simplified-agent-flow} consists of the interview system, and the system that supports the interview designers in testing and distributing the interviews. We begin with the former. As shown in \autoref{fig:simplified-agent-flow}, it comprises the interview guide which structures a given interview (\cref{sec:guide}), and the multi-agent LLM pipeline that generates outputs conditioned on the interview guide  (\cref{sec:agents}). The features that support the testing and distribution of interviews will be covered in \cref{sec:interactive_experience}.

\subsection{Interview guide} \label{sec:guide}
Each interview is defined by an \textit{interview guide} that provides structure in accordance with guidelines for effective interview design, established in social science research \cite{gerson_science_2020, kallio2016systematic, kelly2010qualitative, naz2022development}. The guide organises the interview questions into thematic sections. Each section consists of a set of \textit{main questions} and \textit{probes} (follow-up questions), each informed by \textit{context}. The interview designer's task is to specify and populate an interview guide, for which we provide a customizable template. For example, designers can specify the overall interview focus, the focus of each section, precise question formulations, and question ordering—all designed to support high-quality data collection.

\begin{figure*}[!htb]
    \centering
    \includegraphics[width=0.9\linewidth]{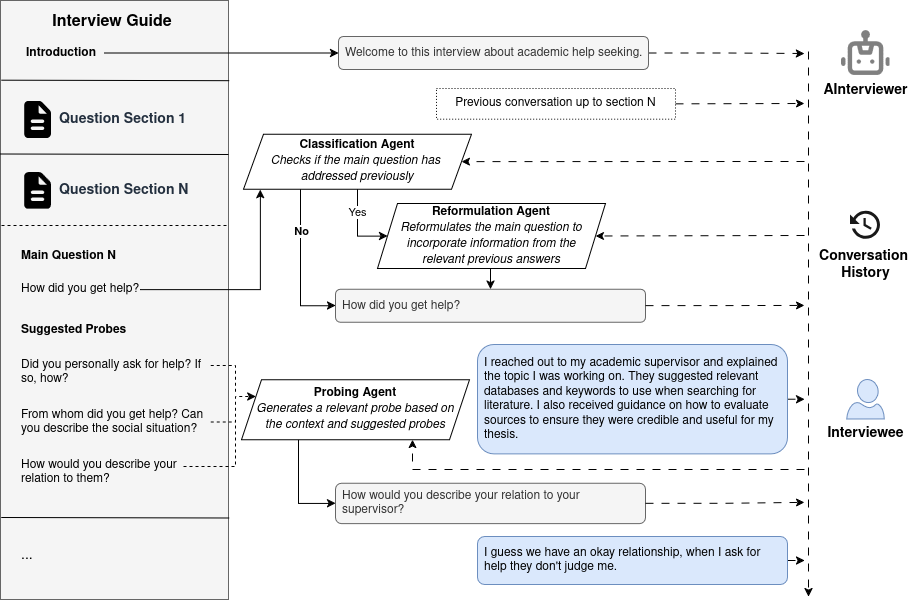}
    \caption{The implementation of the multi-agent AI interview system used in our demonstration. Before generating a main question, a classification agent evaluates whether it has been previously discussed. If that is the case, a reformulating agent is employed to rephrase the main question to fit the conversational context. Once a main question has received a response, the probing agent can generate follow-up questions based on previous conversational context, and any pre-defined probing questions set by researchers.
}
    \label{fig:interview-flow}
\end{figure*}

\subsubsection{Context Fields}
The interview guide allows designers to provide contextual information that models can use to adjust question generation. For example, the \textit{framing} field allows the inclusion of information relevant to the entire interview (e.g., \textit{``This interview is about how people use generative AI'')}, which persists through the thematic shifts. It is also possible to specify context that should only be used within a given thematic section (e.g., \textit{``This section is about the use of generative AI to solve work-related tasks''}).   

\subsubsection{Main questions}
Main questions are pre-defined by the interview designer, and must be asked in each interview to ensure cross-comparability between interviews. To ensure comparability between interviews, the interview designer can set\footnote{All decisions made about the flow of the interview are based on the structure of the interview guide and explicit classification steps. Hence, if it is needed to ensure that main questions are asked verbatim as set in the interview guide, we can simply skip the step where it is reformulated. See \cref{sec:agents} for more details.} them to be asked verbatim as written in the interview guide. Conversely, interview designers can also instruct the system to reformulate pre-defined questions to resolve misunderstandings or to re-introduce previous conversational context, thereby improving data quality \cite{conrad_clarifying_2000, herlihy2024overcoming, sarkar2025conversational}.

\subsubsection{Probes} \label{sec:probes}
We define probes as potential follow-up questions that may be asked depending on the respondent’s answer to a main question. Probes are a core component of semi-structured and unstructured interview methodology \citep{small2022qualitative, robinson2023probing}. Their responsive generation allows the interviewer to support the interviewee’s own narratives and perspectives in relation to the original question, thereby producing in-depth data essential for interpretative inference \cite{gerson_science_2020, lareau2021listening, weiss1995learning, small2022qualitative}. Probing should support interviewees in developing their own answers, and probes should therefore always be relevant to the original response. 

Since probes are context-dependent, the interview guide only provides examples of such questions, and the system retains flexibility to reformulate their wording or generate a more contextually relevant follow-up question based on the unfolding conversation. Alternatively, if no probes are pre-defined, the model will use both its instructions and the context of the interview to generate probes (for more information on probing agents, see \cref{sec:prob_agent}).

\subsection{Multi-agent Pipeline} 
\label{sec:agents}

Interviews are conducted via a multi-agent pipeline, in which each agent is an instantiation of a LLM with its own task. These tasks are specified within system and instruction prompts, which are enriched with dynamic context during the interview. The base model and inference parameters such as temperature can be configured on an per-agent basis. There are three agents: a \textit{probing agent}, a \textit{classification agent}, and a \textit{reformulation agent}. Their roles are described below, and \autoref{fig:interview-flow} shows how they are instructed to interact.%

Our implementation is guided by three core principles: responsiveness, control, and security. These principles led to a strict workflow with clearly divided steps and tasks, distinguishing our approach from other implementations that grant greater responsibility and autonomy to LLM-powered agents, as e.g. in the work by \citet{chopra_conducting_2023}. Furthermore, our system is optimised for using open-source or open-weight LLMs\footnote{See \cref{sec:pilot-eval} for example application with OpenHermes 2.5 7B. We are currently conducting new experiments with gpt-oss-120b, which has improved both the formulation of the probes, and multilingual support.} rather than API-based models, so as to ensure secure data collection, higher control over the model versions and reproducibility, as well as methodological transparency. Our modular architecture facilitates performance evaluations of individual modules, while its adaptability enables re-adjustments to support complex structures beyond this demonstration, making it suitable for many different interview-contexts and situations.

\subsubsection{Probing agent} \label{sec:prob_agent}
The role of this agent is to generate relevant follow-up questions (see \cref{sec:probes}) that, drawing on the interview guide and the respondent's initial answer, elicit further information about the main question being addressed. To do so, the model relies on the conversation transcript, as well as the relevant context sections and suggested probes from the interview guide. Probe generation is further conditioned by comprehensive yet easily modifiable prompts, for example, to specify preferred questioning styles, linguistic register, or instructions for handling sensitive disclosures.

\begin{figure*}[!htb]
    \centering
    \tcbincludegraphics[colframe=gray,width=.65\linewidth]{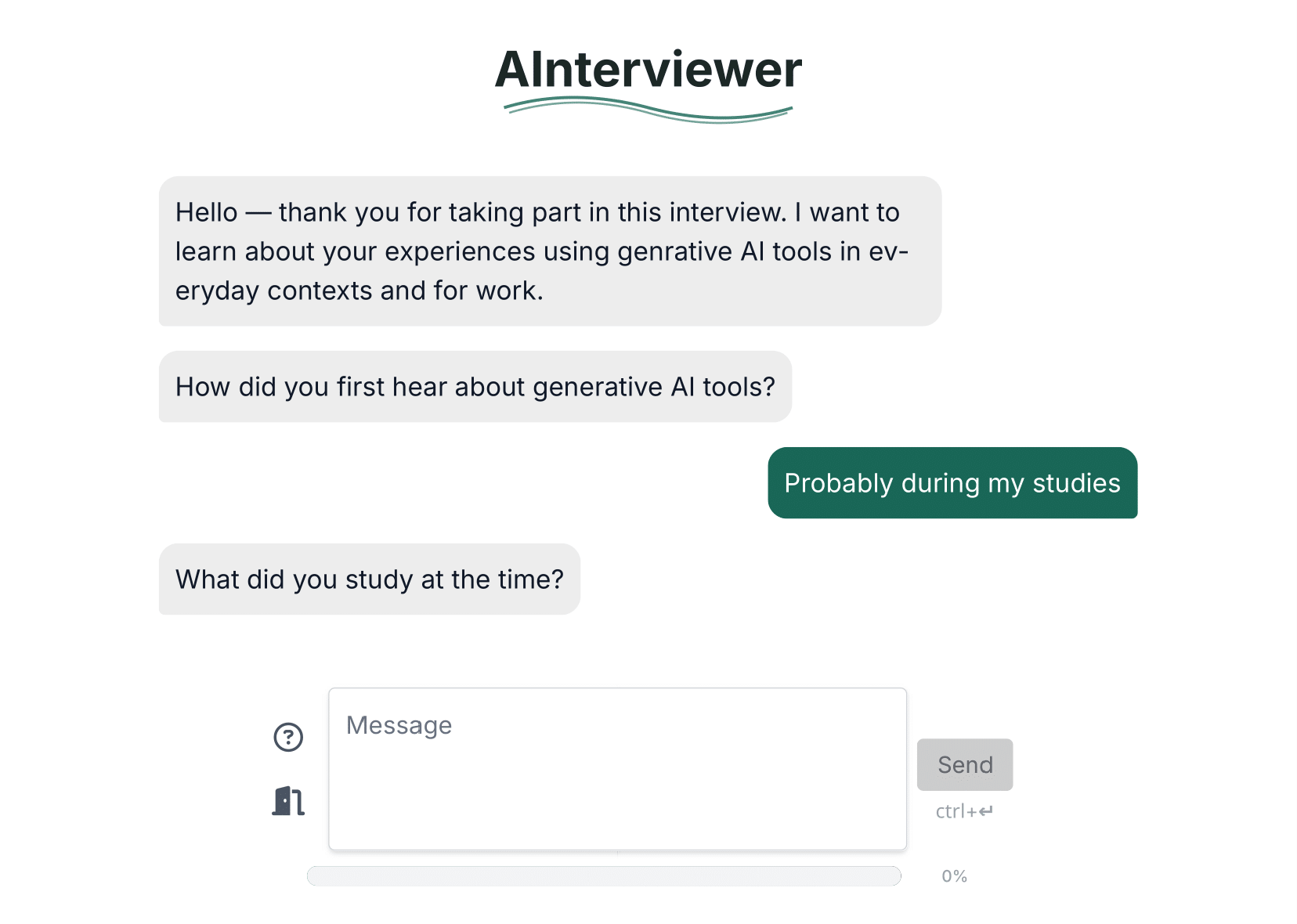}
    \caption{The chat interface that interviewees interact with the AInterviewer.}
    \label{fig:mode}
\end{figure*}

\subsubsection{Reformulating agent}
This agent can be used to reformulate main questions under a set of researcher-specified conditions. For each question, the interview designer can determine whether the reformulation agent should create a smooth conversational transition and/or reformulate a question to account for information previously provided by the respondent. The purpose of the conversational transition is to ensure that the question fits the flow of the conversation without altering its substantive content. For example, if a respondent has already discussed their use of AI at work prior to a hypothetical question about AI and work, the reformulation agent may generate a transition (e.g. `\textit{Let’s talk more about your use of AI at work}'), thereby adding responsive passages without altering the core content of the question. In contrast, reformulating the content of the question involves updating the question itself to incorporate information already provided by the respondent, typically resulting in a more specific question (e.g., `\textit{You said that you mainly use AI to help with writing tasks; can you tell me more about the specific writing tasks you use it for?}') This agent is needed because of our reliance on an otherwise stringent structure imposed by the interview guide. 

\vspace{-0.2em}

\subsubsection{Classification agents}

Classification agents are responsible for altering the flow of the interview including triggering other agents such as the reformulation agent. They do so by evaluating whether specific conditions are met, based on a structured prompt template and relevant context which can include descriptions from the interview guide and transcript of the ongoing interview. The output of the classification agent is a binary value, which triggers the relevant action. We currently implement three classification agents at different points of the interview, to check for the following conditions: 
\begin{itemize}
    \item Whether the question has already been answered in a previous part of the interview, which may trigger the reformulation agent. This is based on the transcript of the current interview, along with the main question description, and is evaluated before each main question is asked. 
    \item Whether the respondent refuses to answer a question, which likewise results in a transition to a new question, and is evaluated before asking each probe. 
    \item Whether a sufficient number of follow-up questions have been asked for the current main question, resulting in a transition to a new question. This is based on the transcript of the current main question, along with the section and main question description, and is evaluated after each probe.
\end{itemize}

\subsection{Interviewee user experience}

Interviewees participate in the interview through a web-based application, as shown in \autoref{fig:mode}. The upper portion of the screen displayed the chat history, while the lower portion features an expanding text box in which interviewees type their responses. The design of the interface has been inspired by messaging applications such as Messenger \cite{meta_messenger}, with a strong emphasis on minimal UI and high-contrast visual elements \cite{lister2020accessible, dong2019minimalist}. The conversation is currently text-based, with its flow mirroring real-life conversations, with dialogue structured around a turn-taking format. Voice-based interviewing interface is in development. 

\section{Workflow for Interview Designers} \label{sec:interactive_experience}

\begin{figure*}[!htb]
    \centering
    \includegraphics[width=0.9\linewidth]{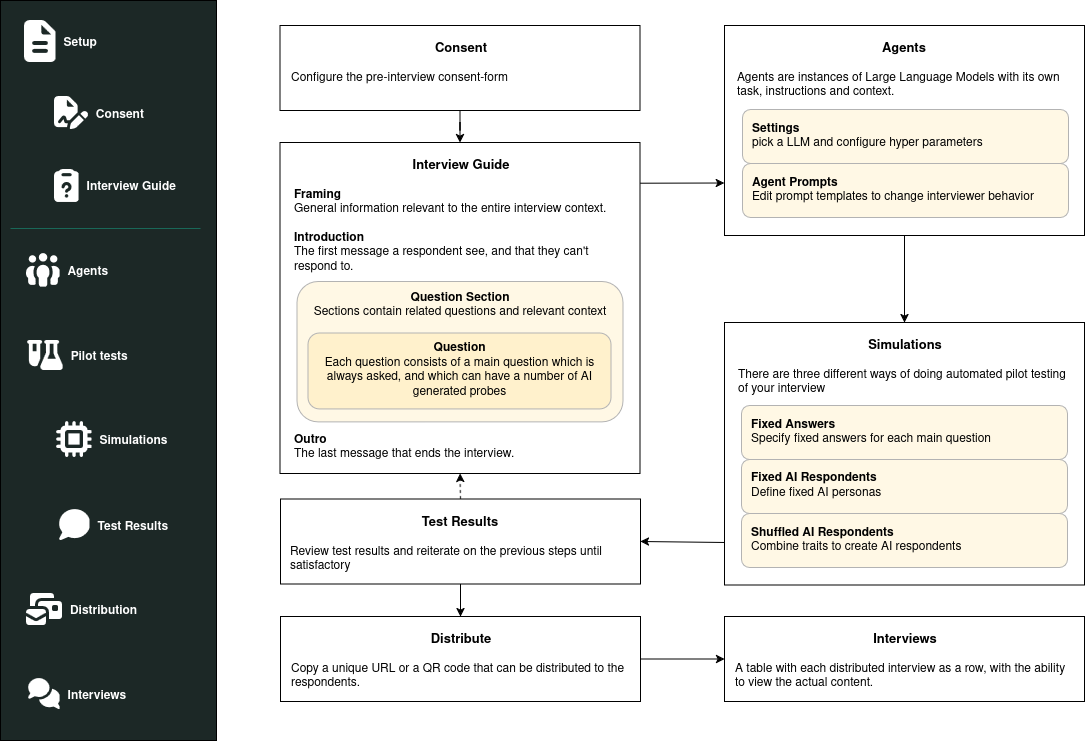}
    \caption{The full workflow for Interview Designers. The setup section (Consent, Interview guide) contains information that must be pre-determined prior to the conduct of the study. The Agents section allows the customising of the multi-agent pipeline, which can be used to pilot prototypes in simulations. Once a study is finalised, it can be easily distributed to participants to partake in AI-led interviews.}
    \label{fig:setup}
\end{figure*}

\textit{AInterviewer} %
can support social science researchers across all stages of interview-based studies, enabling them to easily design, prototype, distribute, and monitor interviews without extensive prior experience in NLP. The full workflow is shown in \autoref{fig:setup}, and we describe the main steps below.

\paragraph{Designing an interview guide}

As discussed in \cref{sec:guide}, interview designers need to define the thematic sections of the interview, formulate the main questions and probes within each section, and determine the overall question order. In practice, interview guides often undergo multiple iterations before deployment. Accordingly, interview designers may proceed to pilot testing (see below) with a preliminary interview guide, rather than aim to produce a fully developed guide at once.

\paragraph{Agent configuration}

This phase involves selecting the LLM models to use for each agent (either via locally hosted models or through third-party API providers) and setting their hyperparameters. In addition, the default agent prompts can be customised to tailor the agents to specific interview objectives and/or specific model.

\paragraph{Pilot testing} This functionality simulates interviews in a controlled environment, enabling the researchers to test their guide and agent configuration before deploying it to real subjects. 
Interviewee responses can be either pre-defined by researchers to test specific conditions, or automatically generated by customizable synthetic agents. The synthetic respondents can be created using pre-defined personas or randomised background characteristics, as shown in \autoref{fig:setup}. During the pilot phase, users can test multiple aspects of the AI interview, including (1) the content of the questions and their formulation, (2) the order of the interview, (3) the performance of the different LLM agents, and (4) the system’s ability to handle high-risk situations. 

\paragraph{Distributing the interview}
Once the interview design is finalised, the researchers can distribute interviews to participants by sharing automatically generated unique URLs or QR codes. The platform also supports the automatic allocation of respondents across different interview versions. For example, controlled interview versions may differ in the wording of specific questions, interview length, or supplementary materials, allowing researchers to test different treatment effects (e.g., differences in participant disclosure based on question ordering).

\paragraph{Data management and sharing}

Researchers can access interview data directly through the Control Panel in its native HTML format or export it in their preferred file format (e.g., CSV). The resulting data files can then be imported into qualitative analysis software such as NVivo \cite{lewis2004nvivo} or into software environments that support quantitative text analysis, such as R. In addition, \textit{AInterviewer} supports sharing the exact system configuration of a given interview study, enabling straightforward replication of the full setup using the accompanying open-source system, from the interview guide, to the precise agent prompts.

\section{System evaluation: pilot study}
\label{sec:pilot-eval}

\textit{AInterviewer} was evaluated in a pilot study with 40 on-site participants. This study is described in detail in separate concurrent work that focuses on the comparative analysis of the quality of data obtained by \textit{AInterviewer} vs trained human interviewers \cite{guven2025comparing}. For completeness, we include a summary of these results with the present technical description of \textit{AInterviewer} system. This study was conducted with OpenHermes-2.5-Mistral-7B\footnote{An open-source model finetuned on Mistral 7B~\cite{jiang2023mistral7b}: \url{https://huggingface.co/teknium/OpenHermes-2.5-Mistral-7B}} as the underlying LLM for the multi-agent pipeline.

\paragraph{Study setup.} The participants\footnote{The participants were university students, recruited through face-to-face outreach on campus and via flyers and posters. The topic of the interviews was university students' help-seeking practices and experiences in the academic setting. The compensation for participation was a voucher for a free lunch at the university canteen. All the interviews were conducted in English. The study was approved by the Institutional Review Board at the University of Copenhagen.} were randomly assigned to be interviewed by either a trained human interviewer or \textit{AInterviewer} via a chat interface. Informed consent was obtained, and participants were aware of the experimental condition (e.g., whether they were participating in a human-led or AI-led interview). %
In total, 946 responses were collected and evaluated for three metrics widely used in qualitative research: exposure (operationalized as length of replies obtained from the participants and number of question answered), relevance (i.e. addressing the topic of the interview) and specificity (providing concrete details of the participants' lived experience). Specificity and relevance were manually coded by the authors of the study on a 5-point scale.

\paragraph{Results.} We compare the two independent groups through a Mann-Whitney U (MWU) test \citep{mann1947test}, followed by a Bonferroni correction \citep{bender2001adjusting} to control for multiple comparisons, as shown in \autoref{tab:mwu-results}.

\begin{table}[t]
    \centering
    \footnotesize
    \begin{tabular}{l c c}
        \toprule
         & U-val & p-val\\
        \midrule
        Response Length & 142.0 & 0.470 \\
        Interview Length & 329.0 & 0.012 \\
        Questions per Interview & 392.0 & 0.000\\
        Mean Specificity & 185.5 & 1.000 \\
        Mean Relevance & 161.0 & 1.000 \\
        \bottomrule
    \end{tabular}
    \caption{Results of the Mann–Whitney U analyses comparing the two independent groups (interviews led by trained human interviewers vs \textit{AInterviewer}) across all outcome variables. For each measure, the table reports the U statistic and the Bonferroni-adjusted p-value.}
    \label{tab:mwu-results}
\end{table}

When assessing response lengths on an individual question basis, we find that human interviewers elicited longer responses (38\% increase), although this effect was not statistically significant ($p_{MWU} = 0.012$, U = 329.0). However, when comparing total response length over whole interviews, the trend reverses, with AI-led interviews eliciting substantially larger corpora ($p_{MWU} = 0.012$, U = 329.0). This effect is explained by the time required to formulate questions between the human and AI interviewers, with AI interviewers outperforming humans by a wide margin.

Regarding response quality, our evaluative metrics indicate no significant differences in specificity and relevance: relevances score were the same (M=4.9). For specificity, human interviews scored higher, but this effect was not statistically significant (M=3.8 vs M=3.5). We interpret these findings as overall promising for AI-led interviewing, although by no means indicating that it can replace expert human interviewers.   %

\section{Conclusion and Future Work}

We present \textit{AInterviewer}, an open-source platform for conducting large-scale AI-led qualitative interviews that is highly customisable and fully transparent in its design. Compared to existing solutions that mostly rely on commercial LLMs accessed via APIs, \textit{AInterviewer} offers a greater control over both the underlying model and the interview process, which is designed to strictly follow an interview guide. Our solution is grounded in best practices of qualitative interview research, and supports the researchers across the entire workflow of developing, testing, and deploying the interviews.

\paragraph{Future work.} Currently \textit{AInterviewer} supports text-based interviews, and future work will focus on incorporating speech functionality to enable more natural interactions between participants and the interview system. We also plan to equip the platform with tools for qualitative and quantitative analysis of the collected interviews.  

\paragraph{Limitations.} While the results of the pilot study summarized in \cref{sec:pilot-eval} are positive, they are preliminary and should not be interpreted as the final verdict on the quality of AI-led interviewing. It is generally a new method for social science research, and its evaluation methodology remains to be established. This includes not only the criteria for comparison to expert human interviewers, but also AI-specific criteria including agent safety, adherence to instructions, linguistic quality of generated questions, etc. Currently, there is no consensus on which constructs and operationalisations should be included in evaluation \citep{wuttke_ai_2024, geiecke_conversations_2024, chopra_conducting_2023}.

\section{Ethical Considerations}

\paragraph{Safety risks for the participants.} Like in other proposed implementations of AI-interview systems, %
the generation process itself remains stochastic. This entails the usual safety problems for all applications of large language models, including unintended outputs, steering the conversation towards unintended topics, or hallucinations when prior conversational content is referenced in long-form conversations \citep{huang2025survey, narayanan-venkit-etal-2024-audit}. There is also the possibility that the interviewees could deliberately %
undermine or `jailbreak' the system during interviews \citep{NEURIPS2024_6d56bc83, shneiderman2020bridging}. These issues apply to all AI-led interview systems, and addressing them will require a field-wide effort. %

\paragraph{Safe interviewing of vulnerable participants.} Interview systems handling responses from %
potentially vulnerable participants %
need to ensure the highest level of safety for them in the interview process. For example, participants may mention something like self-harm, which should be flagged for professional assistance. While the explicit addition of agents responsible for flagging potential safety-critical conversational turns and interviews is planned, the current version of our system is not equipped to handle sensitive information from vulnerable population samples, which makes it appropriate for use only in highly controlled settings. %
Explicit safeguarding of participants during AI-led interviews has also not been implemented in most prior work \citep{wuttke_ai_2024, chopra_conducting_2023, geiecke_conversations_2024, xiao_if_2020, xiao_tell_2020}

\section{Acknowledgements} 

This work was supported by a research grant ([VIL69261]) from Villum Fonden.

\bibliography{custom}

\end{document}